\documentstyle[12pt]{article}
\parskip =\medskipamount
\renewcommand{\arraystretch}{1.2}
\def\be{\begin{equation}}
\def\ee{\end{equation}}
\def\disp{\displaystyle}

\def\scri{\scriptsize}

\newtheorem{conjecture}{Conjecture}

\newcounter{fig}

\def\R{\rm {\sf R\hspace{-3.05mm}I\hspace{2.1mm}}}
\def\Z{\rm {\sf Z\hspace{-3.15mm}Z\hspace{0.6mm}}}

\begin{document}

\begin{titlepage}

\begin{center}
{\LARGE \bf Random Operator Approach for Word \\ Enumeration in Braid Groups

}
\vspace{0.4in}

{\large {\sc Alain Comtet}$^{\dag}$ \hspace{2cm} {\sc Sergei
Nechaev}$^{\ddag\dag}$}
\bigskip

$^{\dag}$ Institut de Physique Nucl\'eaire, Division de Physique
Th\'eorique$^{*}$, \\ 91406 Orsay Cedex, France

$^{\ddag}$ L.D. Landau Institute for Theoretical Physics, \\ 117940, Moscow,
Russia
\end{center}

\vspace{0.4in}

\begin{abstract}
We investigate analytically the problem of enumeration of nonequivalent
primitive words in the braid groups $B_n$ for $n\gg 1$ by analysing the
random word statistics and target space on the basis of the
locally free group approximation. We develop a
"symbolic dynamics" method for exact word enumeration in locally
free groups and bring arguments in support of the conjecture that the number of
very long primitive words in the braid group is not sensitive to the precise
local commutation relations. We consider the connection of these
problems with the conventional random operator theory,
localization phenomena, and statistics of systems with quenched disorder.
Also we discuss the relation of the particular problems of random operator
theory to the theory of modular functions.
\end{abstract}

\vspace{0.2in}

\noindent {\sl Submitted to Nuclear Physics B [Physical Mathematics]}
\vspace{0.2in}

\vspace{0.3in}

\hrule \footnotesize
$^{*}$ Unit\'e de Recherche des Universit\'es Paris XI et Paris VI
associ\'ee au C.N.R.S.

\end{titlepage}

{
\small
\tableofcontents
}

\section{Introduction: Problems and Motivations}

Recent years have been marked by the emergence of more and more
problems related to the consideration of physical
processes on noncommutative groups. In trying to classify
such problems, we distinguish between the following categories in which
the noncommutative origin of phenomena appear with perfect clarity:

1. Problems connected with the spectral properties of the
Harper--Hof\-stad\-ter equation \cite{hofst} dealing with the electron
dynamics on the lattice in a constant magnetic field. We mean primarily 
 the consideration of groups of magnetic translations
\cite{beliss,zabr} and properties of quantum planes \cite{manin}.

2. Problems of classical and quantum chaos on hyperbolic manifolds:
spectral properties of dynamical systems and derivation of trace formulae
\cite{gutzw,terr,bog} as well as construction of probability measures for
random walks on modular groups \cite{chass}.

3. Problems giving rise to application of quantum group theory in physics:
deformations of classical abelian objects such as harmonic oscillators
\cite{gomez} and standard random walks \cite{majid}.

4. Problems of knot theory and statistical topology: construction of
nonabelian topological invariants \cite{jones,kauff}, consideration of
probabilistic behavior of the words on the simplest noncommutative groups
related to topology (such as braid groups) \cite{nech}, statistical
properties of "anyonic" systems \cite{anyon}.

5. Classical problems of random matrix and random
operator\footnote{Following L.A. Pastur, we will distinguish random
matrices and matrix representations of random operators. To the random
operators we attribute the $n\times n$ tables having of order $n$
random entries; if the number of random entries grows faster than $n$ when
$n\to\infty$, we call such a table as a random matrix.} theory and
localization phenomena: determination of Lyapunov exponents for products
of random noncommutative matrices \cite{fuerst_tut,nesi1,nesi2}, study of
the spectral properties and calculation of the density of states of large
random matrices \cite{lifsh,yan}.

Certainly, such a division of problems into these categories is
very speculative and reflects to a marked degree the authors' personal
point of view. However, we believe that the enumerated items reflect, at
least partially, the currently growing interest in theoretical physics of
the ideas of noncommutative analysis.

Let us stress that we do not touch upon the pure mathematical aspects of
noncommutative analysis in this paper and the problems discussed in the
present work mainly concern the points 4 and 5 of the list above.

It is widely considered that a new fresh stream in topology was brought
about by the recognition of the fact that there exists a deep relation
between the Temperley--Lieb algebra and the Hecke algebra representation of
the braid group. This fact resulted in the remarkable geometrical analogy
between the Yang--Baxter equations, appearing as a necessary condition of
the transfer matrix commutativity in the theory of integrable systems on
the one hand, and the Reidemeister moves, used in the knot invariant
construction on the other hand. We can mention several nice reviews
\cite{jones,likor,wad} and books \cite{collect,kauff} regarding the
construction of new knot and link invariants in terms of integrable 2D
statistical models, as well as their relation with different matrix and
tensor representations of some noncommutative groups.

Besides the traditional topological issues such as construction of
topological invariants, investigation of homotopic classes and fibre
bundles, we consider a set of adjacent but much less studied problems lying
at the border between statistical physics, probability theory and
topology. First of all, we should mention problems related to the
so-called "knot
entropy" calculation. Most generally this set of problems can be formulated
as follows. Take the lattice ${\Z}^3$ embedded in the space ${\R}^3$. Let
$\Omega_N$ be the ensemble of all possible closed non-self-intersecting
$N$--step loops with one common fixed point on ${\Z}^3$; by $\omega$ we
denote the particular trajectory configuration. The question is: what is
the probability ${\cal P}_N$ of the trajectory $\omega\in\Omega_N$
belonging to some specific homotopy class. In \cite{nech} it has been
shown that many non-trivial properties of the limit behavior of knot
statistics can be explained in the context of the limit behavior of random
walks over the elements of some nonabelian (hyperbolic) group related to
the braid representation of knots.

In the context of "topologically--probabilistic" consideration, the
problems dealing with the limit distributions of noncommutative random
walks were not discussed practically except very few specific
cases \cite{nesi1,nesi2,khne,nesem}. In particular, in these works it has
been shown that statistics of random walks with a fixed topological state
with respect to the regular arrays of obstacles on the plane can be mapped
to random walks on the free group $\Gamma_2$ which has the topology of a
simply connected tree. The analytic construction of nonabelian topological
invariants for trajectories on a double punctured plane as well as the
statistics of the simplest nontrivial random braid $B_3$, were  discussed
briefly in \cite{never,mona}.

A preliminary analytical and numerical study of the statistics
of random walks (Markov chains) on braid and so-called "locally free"
groups \footnote{This notation has been introduced by A.M. Vershik in
\cite{nevegr}.} (see definition below) was recently undertaken in works
\cite{nevegr,jean}. In the case of the braid group, the rather complicated
group structure prevents us from applying the simple geometrical pictures
of the free group $\Gamma_2$ (see \cite{kesten}). Nevertheless the
problem of the limit distribution for random walks on $B_n$ can be reduced
to the problem of a random walk on some graph \cite{nevegr,jean}. In case
of the group $B_3$ we were able to construct this graph explicitly, whereas
for the group $B_n$ ($n\ge 4$) we gave only an upper estimate for the
limit distribution of random walks analysing statistics of Markov chains on
"locally free groups".

The consideration of problems dealing with the limit distributions of Markov
chains on braid groups $B_n$ requires examination of the "target space" of this
group, i.e., the space where the random walk takes place. The structure of the
target space is uniquely determined by the group relations and in the
general case of the group $B_n$ ($n\ge 4$), is still unknown.

In the present work we study the target space of
the braid group $B_n$ when $n\gg 1$, trying to develop a new "statistical
approach" for words enumeration in this group.

We should stress that our presentation offers a mathematical analysis
which is far from rigorous, and ideas expressed here are mainly
supported by numerical simulations. Moreover, we skip here some important
but hard questions, like the problem of "words identity" in the braid
group (deep advances concerning this subject can be found in recent
work \cite{birm_prep}). Our aim is to describe a constructive algorithm
which, of course, has to be be justified and verified later.

The structure of the paper is as follows: in the next section we give
some necessary definitions concerning braid and "locally free" groups and
describe the model under consideration; Section 3 is devoted to developing
a "symbolic dynamics" method for words enumeration in  the locally free
group ${\cal LF}_n$ (for $n\gg 1$); the target space of the braid group is
studied in Section 4 by means of a statistical approach based on the concept
of "locally free group with errors". In this section we discuss also some
additional links between this problems and the conventional random matrix
theory, localization phenomena and statistics of systems with "quenched"
disorder; we present here some speculations dealing with the possible
relation of particular problems of random matrix theory to the theory of
Dedekind function and modular groups.

\section{Basic Definitions and Statistical Model}

We first recall some points concerning the definitions of braid and
"locally free" groups.
\bigskip

\noindent{\sc Braid Group}. The braid group $B_n$ of "$n$--strings" has
$n-1$ generators $\{\sigma_1,\sigma_2,\ldots,\sigma_{n-1}\}$ with the
following commutation relations:
\be \label{2:1}
\begin{array}{ll}
\sigma_i\sigma_{i+1}\sigma_i = \sigma_{i+1}\sigma_i\sigma_{i+1}
& \qquad (1\le i<n-1) \\
\sigma_i\sigma_j=\sigma_j\sigma_i & \qquad (|i-j|\ge 2) \\
\sigma_i\sigma_i^{-1}=\sigma_i^{-1}\sigma_i=e &
\end{array}
\ee

\noindent Let us mention that:

\noindent -- A word written in terms of "letters" --- generators from
the set $\{\sigma_1,\ldots, \sigma_{n-1},\sigma_1^{-1},\ldots,\break
\sigma_{n-1}^{-1}\}$ gives a particular {\it braid}. Schematically the
generators $\sigma_i$ and $\sigma_i^{-1}$ may be represented as follows:

\bigskip

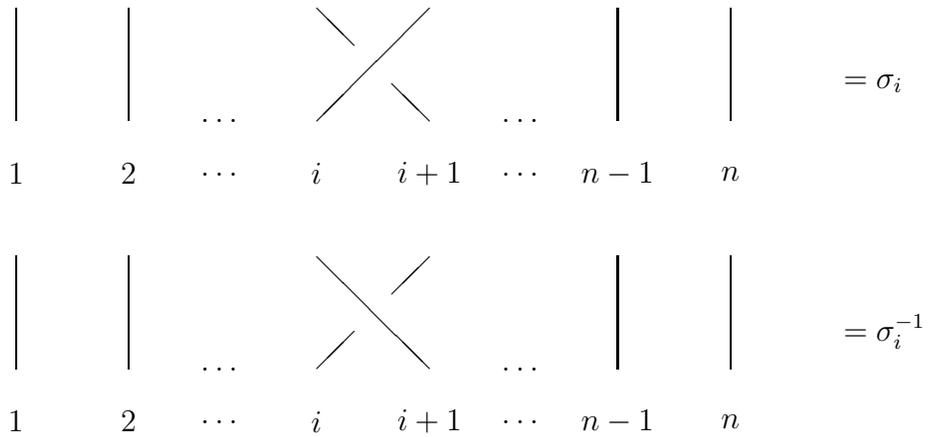
\unitlength=1.00mm
\special{em:linewidth 0.5pt}
\linethickness{0.5pt}
\hspace{2.5cm}
\begin{picture}(100.00,50.00)
\put(5.00,50.00){\line(0,-1){15.00}}
\put(30.00,50.00){\line(1,-1){5.00}}
\put(40.00,40.00){\line(1,-1){5.00}}
\put(45.00,50.00){\line(-1,-1){15.00}}
\put(-10.00,50.00){\line(0,-1){15.00}}
\put(70.00,50.00){\line(0,-1){15.00}}
\put(85.00,50.00){\line(0,-1){15.00}}
\put(5.00,17.00){\line(0,-1){15.00}}
\put(-10.00,17.00){\line(0,-1){15.00}}
\put(70.00,17.00){\line(0,-1){15.00}}
\put(85.00,17.00){\line(0,-1){15.00}}
\put(30.00,17.00){\line(1,-1){15.00}}
\put(45.00,17.00){\line(-1,-1){5.00}}
\put(35.00,7.00){\line(-1,-1){5.00}}
\put(-10.00,28.00){\makebox(0,0)[cc]{$1$}}
\put(5.00,28.00){\makebox(0,0)[cc]{$2$}}
\put(17.00,35.00){\makebox(0,0)[cc]{$\ldots$}}
\put(17.00,28.00){\makebox(0,0)[cc]{$\ldots$}}
\put(30.00,28.00){\makebox(0,0)[cc]{$i$}}
\put(45.00,28.00){\makebox(0,0)[cc]{$i+1$}}
\put(57.00,35.00){\makebox(0,0)[cc]{$\ldots$}}
\put(57.00,28.00){\makebox(0,0)[cc]{$\ldots$}}
\put(70.00,28.00){\makebox(0,0)[cc]{$n-1$}}
\put(85.00,28.00){\makebox(0,0)[cc]{$n$}}
\put(100.00,40.00){\makebox(0,0)[ll]{$\large =\sigma_i$}}
\put(-10.00,-5.00){\makebox(0,0)[cc]{$1$}}
\put(5.00,-5.00){\makebox(0,0)[cc]{$2$}}
\put(17.00,2.00){\makebox(0,0)[cc]{$\ldots$}}
\put(17.00,-5.00){\makebox(0,0)[cc]{$\ldots$}}
\put(30.00,-5.00){\makebox(0,0)[cc]{$i$}}
\put(45.00,-5.00){\makebox(0,0)[cc]{$i+1$}}
\put(57.00,2.00){\makebox(0,0)[cc]{$\ldots$}}
\put(57.00,-5.00){\makebox(0,0)[cc]{$\ldots$}}
\put(70.00,-5.00){\makebox(0,0)[cc]{$n-1$}}
\put(85.00,-5.00){\makebox(0,0)[cc]{$n$}}
\put(100.00,7.00){\makebox(0,0)[ll]{$\large =\sigma_i^{-1}$}}
\end{picture}

\vspace{0.3in}

\noindent -- The {\it length} of the braid is the total number of
letters used, while the {\it minimal irreducible length} hereafter
referred to as "primitive word" is the shortest noncontractible length of a
particular braid remaining after all possible group relations
Eq.(\ref{2:1}) are applied. Diagramatically, the braid can be represented
as a set of crossed strings going from the top to the bottom  after gluing
the braid generators.

\noindent -- The closed braid appears after gluing the "upper" and the
"lower" free ends of the braid on the cylinder.

\noindent -- Any braid corresponds to some knot or link. So, there is a
possibility to use the braid group representation for the
construction of topological invariants of knots and links, but the
correspondence between braids and knots is not mutually single valued and
each knot or link can be represented by an infinite series of different
braids.
\bigskip

\noindent {\sc Locally Free Group}. The group ${\cal LF}_n(d)$ is called
{\it locally free} if the generators, $\{f_1,\ldots,f_{n-1}\}$ obey the
following commutation relations:
\begin{itemize}
\item[(a)] Each pair $(f_j, f_k)$ generates the free subgroup of the group
${\cal LF}_n(d)$ if $|j-k|<d$;
\item[(b)] $f_j f_k=f_k f_j$ for  $|j-k|\ge d$
\end{itemize}
We will be concerned mostly with the case $d=2$ for which we define
${\cal LF}_n(2)\equiv {\cal LF}_n$.

\noindent -- The {\it length} of the word written in terms of letters
$\{f_1,\ldots,f_{n-1} ,f_1^{-1},\ldots, f_{n-1}^{-1}\}$ is the total number
of  generators used, and the "primitive word" is the shortest
noncontractible length of a particular word after applying all relations of
the group ${\cal LF}_n(2)$ (compare to the case of the braid group). The
graphical representation of generators $g_i$ and $g_i^{-1}$ is also rather
similar to that of braid group:

\bigskip
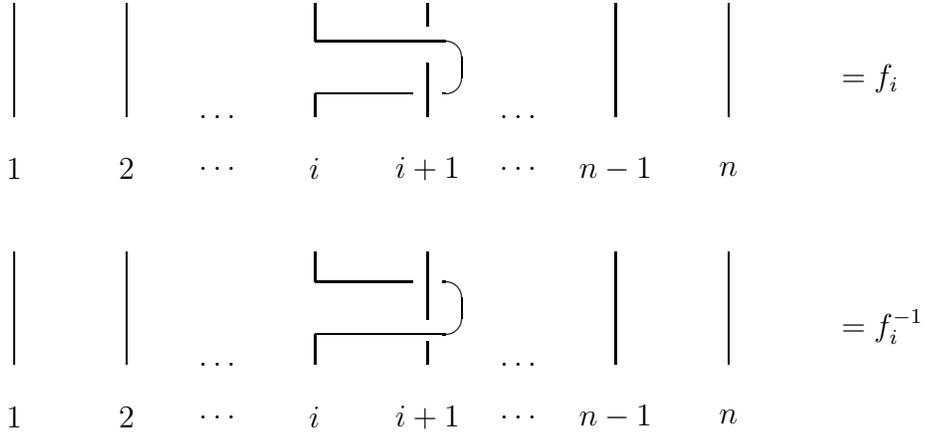
\unitlength=1.00mm
\special{em:linewidth 0.5pt}
\linethickness{0.5pt}
\hspace{2.5cm}
\begin{picture}(100.00,50.00)
\put(5.00,50.00){\line(0,-1){15.00}}
\put(-10.00,50.00){\line(0,-1){15.00}}
\put(70.00,50.00){\line(0,-1){15.00}}
\put(85.00,50.00){\line(0,-1){15.00}}
\put(5.00,17.00){\line(0,-1){15.00}}
\put(-10.00,17.00){\line(0,-1){15.00}}
\put(70.00,17.00){\line(0,-1){15.00}}
\put(85.00,17.00){\line(0,-1){15.00}}
\put(-10.00,28.00){\makebox(0,0)[cc]{$1$}}
\put(5.00,28.00){\makebox(0,0)[cc]{$2$}}
\put(17.00,35.00){\makebox(0,0)[cc]{$\ldots$}}
\put(17.00,28.00){\makebox(0,0)[cc]{$\ldots$}}
\put(30.00,28.00){\makebox(0,0)[cc]{$i$}}
\put(45.00,28.00){\makebox(0,0)[cc]{$i+1$}}
\put(57.00,35.00){\makebox(0,0)[cc]{$\ldots$}}
\put(57.00,28.00){\makebox(0,0)[cc]{$\ldots$}}
\put(70.00,28.00){\makebox(0,0)[cc]{$n-1$}}
\put(85.00,28.00){\makebox(0,0)[cc]{$n$}}
\put(100.00,40.00){\makebox(0,0)[ll]{$\large =f_i$}}
\put(-10.00,-5.00){\makebox(0,0)[cc]{$1$}}
\put(5.00,-5.00){\makebox(0,0)[cc]{$2$}}
\put(17.00,2.00){\makebox(0,0)[cc]{$\ldots$}}
\put(17.00,-5.00){\makebox(0,0)[cc]{$\ldots$}}
\put(30.00,-5.00){\makebox(0,0)[cc]{$i$}}
\put(45.00,-5.00){\makebox(0,0)[cc]{$i+1$}}
\put(57.00,2.00){\makebox(0,0)[cc]{$\ldots$}}
\put(57.00,-5.00){\makebox(0,0)[cc]{$\ldots$}}
\put(70.00,-5.00){\makebox(0,0)[cc]{$n-1$}}
\put(85.00,-5.00){\makebox(0,0)[cc]{$n$}}
\put(100.00,7.00){\makebox(0,0)[ll]{$\large =f_i^{-1}$}}
\put(30.00,38.00){\line(1,0){13.00}}
\put(30.00,50.00){\line(0,-1){5.00}}
\put(30.00,38.00){\line(0,-1){3.00}}
\put(45.00,50.00){\line(0,-1){3.00}}
\put(45.00,42.00){\line(0,-1){7.00}}
\put(30.00,13.00){\line(1,0){13.00}}
\put(30.00,17.00){\line(0,-1){4.00}}
\put(30.00,6.00){\line(0,-1){4.00}}
\put(45.00,17.00){\line(0,-1){9.00}}
\put(45.00,5.00){\line(0,-1){3.00}}
\put(47.00,41.48){\oval(5.06,6.97)[r]}
\put(47.00,45.00){\line(-1,0){16.95}}
\put(47.00,9.45){\oval(5.06,6.97)[r]}
\put(47.00,6.00){\line(-1,0){16.90}}
\end{picture}

\vspace{0.3in}

It is easy to understand that the following geometrical identity is valid:

\bigskip
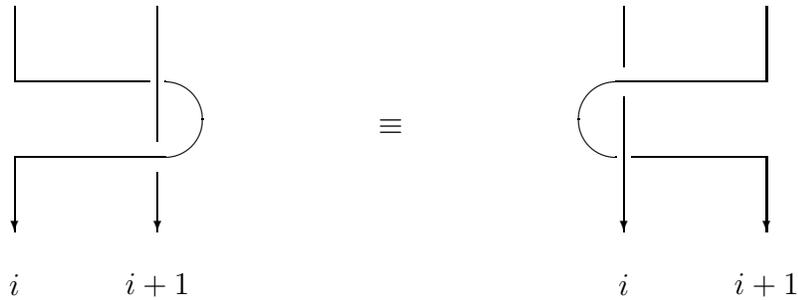
\unitlength=1.00mm
\special{em:linewidth 0.4pt}
\linethickness{0.4pt}
\hspace{1cm}
\begin{picture}(110.00,40.00)
\put(110.00,40.00){\line(0,-1){10.00}}
\put(110.00,20.00){\vector(0,-1){10.00}}
\put(10.00,40.00){\line(0,-1){10.00}}
\put(10.00,20.00){\vector(0,-1){10.00}}
\put(29.00,40.00){\line(0,-1){18.00}}
\put(29.00,18.00){\vector(0,-1){8.00}}
\put(91.00,40.00){\line(0,-1){8.00}}
\put(91.00,28.00){\vector(0,-1){18.00}}
\put(60.00,24.00){\makebox(0,0)[cc]{$\Large \equiv$}}
\put(90.00,25.00){\oval(10.00,10.00)[l]}
\put(110.00,30.00){\line(-1,0){20.00}}
\put(30.00,25.00){\oval(10.00,10.00)[r]}
\put(92.00,20.00){\line(1,0){18.00}}
\put(30.00,20.00){\line(-1,0){20.00}}
\put(10.00,30.00){\line(1,0){18.00}}
\put(10.00,3.00){\makebox(0,0)[cc]{$i$}}
\put(29.00,3.00){\makebox(0,0)[cc]{$i+1$}}
\put(91.00,3.00){\makebox(0,0)[cc]{$i$}}
\put(110.00,3.00){\makebox(0,0)[cc]{$i+1$}}
\end{picture}

\noindent hence, it is unnecessary to distinguish between "left" and
"right" operators $f_i$.

It can be seen that the only difference between the braid and locally
free groups consists in the elimination of the Yang-Baxter relations (first
line in Eq.(\ref{2:1})).
\bigskip

\noindent {\sc Statistical Model}. Our aim is to calculate a specific
"partition function", $V_n(\mu,d)$, giving the number of all
nonequivalent primitive words of length $\mu$ in the groups ${\cal
LF}_{n+1}(d)$ and $B_{n+1}$ for $n\gg 1$.
\bigskip

\noindent {\bf Remark}. To have a geometrical picture of the group ${\cal
LF}_{n+1}$ let us describe the recursion procedure of raising the graph
(the "target space") associated with this group.

Take the {\it free} group $\Gamma_n$ with generators $\{\tilde{f}_1,\ldots,
\tilde{f}_n\}$ where all $\tilde{f}_i$ ($1\le i\le n$) do not commute. It
is well known that the group $\Gamma_n$ has the structure of a $2n$-branching
Cayley tree, $C(\Gamma_n)$---see Fig.\ref{fig:1}a---where the number of
distinct primitive words of length $\mu$ is equal to the number
$\tilde{V}_n(\mu)$ of vertices of the tree $C(\Gamma_n)$ lying at a
distance of $\mu$ steps from the origin:
\be \label{2:vfree}
\tilde{V}_n(\mu)=2n(2n-1)^{\mu-1}
\ee
The graph $C({\cal LF}_{n+1})$ corresponding to the group ${\cal LF}_{n+1}$
can be constructed from the graph $C(\Gamma_n)$ in accordance with the
following recursion procedure:
\begin{itemize}
\item[(a)] Take the root vertex of the graph $C(\Gamma_n)$ and consider all
vertices on the distance $\mu=2$ from it. Identify those vertices which
correspond to the equivalent words in the group ${\cal LF}_{n+1}$. (See
example in Fig.\ref{fig:1}b).
\item[(b)] Repeat this procedure taking all vertices at the distance
$\mu=(1,2,\ldots)$ and "gluing" them at the distance $\mu+2$ according to the
definition of the locally free group.
\end{itemize}
By means of the procedure described, we raise a graph ("target space")
corresponding to the locally free group ${\cal LF}_{n+1}$. Now our main
problem can be reformulated as follows: how many {\it distinct} vertices
has the graph $C({\cal LF}_{n+1})$ at a distance of $\mu$ steps from
the origin (for $n\gg 1$).

In the next Section we give an exact answer to that question, which we use, in
turn, as the basis for consideration of the much trickier case of the
braid group.

It is worthwhile to mention that the graph $C({\cal E}_{n+1})$ of the
complete commutative group ${\cal E}_{n+1}$ (all generators of
${\cal E}_{n+1}$ commute with each others) has the topology of the lattice
embedded in ${\R}^{2n}$ (see Fig.\ref{fig:1}c). Hence, the number of
nonequivalent words $V^{\rm comm}_n(\mu)$ of length $\mu$ can be roughly
estimated as the number of lattice points lying on the surface of the
$2n$--dimensional sphere, i.e.
\be \label{2:vcomm}
V^{\rm comm}_n(\mu)\simeq {\rm const}\; \mu^{2n}
\ee

Comparing (\ref{2:vfree}) and (\ref{2:vcomm}) we get
\be \label{2:comp}
\left\{\begin{array}{lcl}
\disp \lim_{\scri \shortstack{\\ $\mu\to\infty$ \\ $n={\rm const}\gg 1$}}
\frac{1}{\mu} \ln \tilde{V}_n(\mu)&\simeq &\ln (2n-1) >0 \medskip \\
\disp
\lim_{\scri \shortstack{\\ $\mu\to\infty$ \\ $n={\rm const}\gg 1$}}
\frac{1}{\mu}\ln V^{\rm comm}_n(\mu)&=&0 \end{array} \right.  \ee

Naively we could expect that the behavior
$$
\disp \lim_{\scri \shortstack{\\
$\mu\to\infty$ \\ $n={\rm const}\gg 1$}}
\frac{1}{\mu}\ln V_n(\mu)=0
$$
for "locally free" and braid groups remains unchanged (i.e. is the same as
in the completely commutative case) because for $n\gg 1$ we have of order
$\sim n^2$ commutative relations and only of order $\sim n$ noncommutative
ones. However, it is proven for ${\cal LF}_n$ and conjectured for $B_n$
that
$$
\disp \lim_{\scri \shortstack{\\
$\mu\to\infty$ \\ $n={\rm const}\gg 1$}}
\frac{1}{\mu}\ln V_n(\mu)={\rm const}>0
$$
which clearly reflects the hyperbolic character of these groups.

\section{Exact Words Enumeration in Locally Free Groups}

We derive here an explicit expression for the number $V_n(\mu,d)$ of all
nonequivalent primitive words of length $\mu$ in the group ${\cal
LF}_{n+1}(d)$ (when $d=2$ and $n\gg 1$) on the basis of the so-called
"normal order" representation of words proposed by A.M. Vershik in
\cite{versh2} and developed in \cite{nevegr,jean} which is reminiscent of
the enumeration of "partially commutative monoids" known in combinatorics
\cite{monoids}.

\subsection{"Normal Order" Representation of Words}

Let us represent each primitive word $W_p$ of length $\mu$ in the group
${\cal LF}_{n+1}(d)$ in the {\bf normal order} similar to the so-called
"symbolic dynamics" appearing in the context  of chaotic systems (see, for
instance, \cite{bog})
\be \label{2:norm}
W_p=\left(f_{\alpha_1}\right)^{m_1}\left(f_{\alpha_2}\right)^{m_2}\ldots
\left(f_{\alpha_s}\right)^{m_s}
\ee
where $\sum_{i=1}^s |m_i|=\mu\; (m_i\neq 0\; \forall\; i;\; 1\le s\le
\mu$) and the sequence of generators $f_{\alpha_i}$ in Eq.(\ref{2:norm})
{\bf for all distinct} $f_{{\alpha}_i}$ satisfies the following local rules
\cite{nevegr}:
\begin{itemize}
\item[(i)] If $f_{\alpha_i}=f_1$, then $f_{\alpha_{i+1}}\in
\left\{f_2,f_3,\ldots f_n\right\}$;
\item[(ii)] If $f_{\alpha_i}=f_k$ ($1<k\le n-1$), then $f_{\alpha_{i+1}}\in
\left\{f_{k-d+1},\dots,f_{k-1},f_{k+1},\ldots f_n\right\}$;
\item[(iii)] If $f_{\alpha_i}=f_n$, then $f_{\alpha_{i+1}}\in
\left\{f_{k-d+1},\ldots, f_{n-1}\right\}$.
\end{itemize}
These local rules could be represented diagramatically as follows:

\bigskip
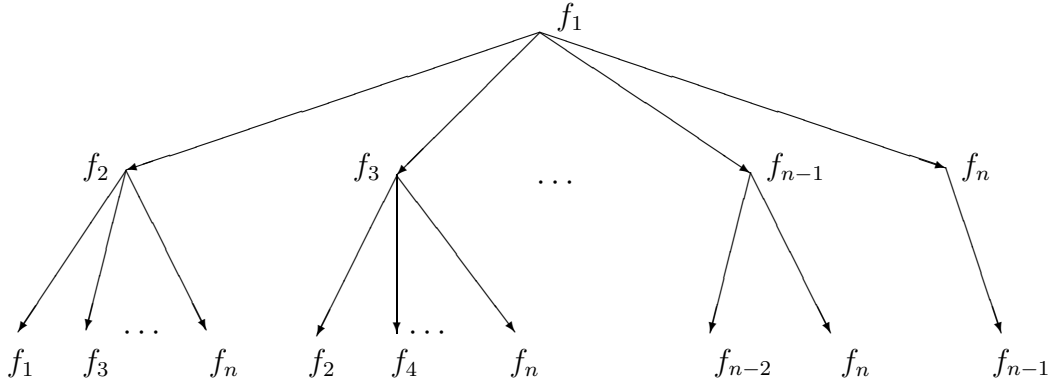
\unitlength=1.00mm
\special{em:linewidth 0.4pt}
\linethickness{0.4pt}
\begin{picture}(138.00,47.00)
\put(79.00,47.00){\makebox(0,0)[cc]{$f_1$}}
\put(16.00,27.00){\makebox(0,0)[cc]{$f_2$}}
\put(52.00,27.00){\makebox(0,0)[cc]{$f_3$}}
\put(133.00,27.00){\makebox(0,0)[cc]{$f_n$}}
\put(6.00,1.00){\makebox(0,0)[cc]{$f_1$}}
\put(16.00,1.00){\makebox(0,0)[cc]{$f_3$}}
\put(33.00,1.00){\makebox(0,0)[cc]{$f_n$}}
\put(46.00,1.00){\makebox(0,0)[cc]{$f_2$}}
\put(57.00,1.00){\makebox(0,0)[cc]{$f_4$}}
\put(73.00,1.00){\makebox(0,0)[cc]{$f_n$}}
\put(77.00,25.00){\makebox(0,0)[cc]{$\cdots$}}
\put(22.00,5.00){\makebox(0,0)[cc]{$\cdots$}}
\put(60.00,5.00){\makebox(0,0)[cc]{$\cdots$}}
\put(102.00,1.00){\makebox(0,0)[cc]{$f_{n-2}$}}
\put(117.00,1.00){\makebox(0,0)[cc]{$f_n$}}
\put(139.00,1.00){\makebox(0,0)[cc]{$f_{n-1}$}}
\put(109.00,27.00){\makebox(0,0)[cc]{$f_{n-1}$}}
\put(75.00,45.00){\vector(-3,-1){55.00}}
\put(20.00,26.67){\vector(-2,-3){14.33}}
\put(20.00,26.67){\vector(-1,-4){5.33}}
\put(20.00,26.67){\vector(1,-2){10.67}}
\put(75.00,45.00){\vector(-1,-1){19.00}}
\put(56.00,26.00){\vector(-1,-2){10.67}}
\put(56.00,26.00){\vector(0,-1){21.00}}
\put(56.00,26.00){\vector(3,-4){15.67}}
\put(75.00,45.00){\vector(3,-2){28.00}}
\put(103.00,26.33){\vector(-1,-4){5.33}}
\put(103.00,26.33){\vector(1,-2){10.67}}
\put(75.00,45.00){\vector(3,-1){54.00}}
\put(129.00,27.00){\vector(1,-3){7.33}}
\end{picture}

\vspace{0.3in}

The rules (i)--(iii) give the prescription how to encode and enumerate all
distinct primitive words in the group ${\cal LF}_{n+1}(d)$. If the
sequence of generators in the primitive word $W_p$ does not satisfy the
rules (i)-(iii), we commute the generators in the word $W_p$ until the
normal order is restored.  Hence, the normal order representation enables
one to give the unique coding of all nonequivalent primitive words in the
group ${\cal LF}_{n+1}(d)$.
\medskip

\noindent {\bf Example 1}. Take an arbitrary primitive word of length
$\mu=10$ in the group ${\cal LF}_{8+1}(2)$:
\be \label{word1}
\begin{array}{ccl}
W_p & = & f_5^{-1}\, f_3\, f_8\, f_1^{-1}\, f_2\, f_4\, f_8\, f_8\, f_4\,
f_7 \\
& \equiv & (f_5)^{-1}\, (f_3)\, (f_8)\, (f_1)^{-1}\, (f_2)\, (f_4)\,
(f_8)^2\, (f_4)\, (f_7)
\end{array}
\ee
To represent the word $W_p$ in the "normal order" we have to push all
generators with smaller indices to the left when it is allowed by the
commutation relations of the locally free group ${\cal LF}_9(2)$. We get:
\be \label{word2}
W_p\,=\,(f_1)^{-1}\, (f_3)^1\, (f_2)^1\, (f_5)^{-1}\, (f_4)^2\, (f_8)^3\,
(f_7)^1
\ee
(the "normal order" for this word is the sequence of used generators:
$\{1,3,2,5,4,8,7\}$).

To compute the number of different primitive words of length $\mu=10$ with
the same normal order as in Eq.(\ref{word2}), we have to sum up all the
words like
\be \label{word3}
W_p\,=\,(f_1)^{m_1}\, (f_3)^{m_2}\, (f_2)^{m_3}\, (f_5)^{m_4}\,
(f_4)^{m_5}\,(f_8)^{m_6}\, (f_7)^{m_7}
\ee
under the condition $\sum_{i=1}^7 |m_i|=10;\; m_i\neq 0\; \forall\; m_i\in
[1,7]$.
\bigskip

The calculation of the number of distinct primitive words, $V_n(\mu)$, of
the given length $\mu$ is now rather straightforward:
\be \label{2:trace}
V_n(\mu,d)=\sum_{s=1}^{\mu} R_n(s,d)\mathop{{\sum}'}_{\{m_1,\ldots,m_s\}}
\Delta\left[\sum_{i=1}^s |m_i|-\mu\right]
\ee
where: \\
$\bullet$ $R_n(s,d)$ is the number of all distinct sequences of $s$
generators taken from the set $\{f_1,\ldots,f_n\}$ and satisfying the local
rules (i)-(iii) \\
$\bullet$ the second sum gives the number of all possible representations
of the primitive path of length $\mu$ {\bf for the fixed sequence of
generators}---(see the example above); "prime" means that the sum runs over
all $m_i\neq 0$ for $1\le i\le s$; $\Delta$ is the Kronecker function:
$\Delta=1$ for $x=0$ and $\Delta=0$ for $x\neq 1$ \\
$\bullet$ special attention should be paid to the sequences built on the
basis of one generator only, i.e. for primitive words of type
$W_p=(f_k)^{\mu}$ $\forall k\in [1,n]$ (see definition of $R_n(s,d)$
below).

To get the partition function $R_n(s,d)$ let us mention that the local rules
(i)-(iii) define a generalized Markov chain with the states given by the
$n\times n$ "incidence" matrix $\widehat{M}_n(d)$, the rows and columns of
which correspond to the generators $f_1,\ldots,f_n$ as it is shown below:

\be \label{matrix}
\widehat{M}_n(d)=
{
\renewcommand{\arraystretch}{0}
\begin{tabular}{|c||c|c|c|c|c|c|c|} \hline
\strut \rule{0pt}{15pt} & $f_1$ & $f_2$ & $f_3$ & $f_4$ & $\ldots$ &
$f_{n-1}$ & $f_n$ \\ \hline \rule{0pt}{2pt} & & & & & & & \\ \hline
\strut \rule{0pt}{15pt} $f_1$ & 0 & 1 & 1 & 1 & $\ldots$ & 1 & 1 \\ \hline
\strut \rule{0pt}{15pt} $f_2$ & 1 & 0 & 1 & 1 & $\ldots$ & 1 & 1 \\ \hline
\strut \rule{0pt}{15pt} $f_3$ & 0 & 1 & 0 & 1 & $\ldots$ & 1 & 1 \\ \hline
\strut \rule{0pt}{15pt} $f_4$ & 0 & 0 & 1 & 0 & $\ldots$ & 1 & 1 \\ \hline
\strut $\vdots$ & $\vdots$ & $\vdots$ & $\vdots$ & $\ddots$ &
$\ddots$ & $\vdots$ & $\vdots$ \\ \hline
\strut \rule{0pt}{15pt} $f_{n-1}$ & 0 & 0 & 0 & 0 & $\ddots$ & 0 & 1 \\
\hline
\strut \rule{0pt}{15pt} $f_n$ & 0 & 0 & 0 & 0 &  & 1 & 0 \\ \hline
\end{tabular}
}
\ee
\medskip

\noindent The matrix $\widehat{M}_n(d)$ has a rather simple structure:
above the diagonal we put everywhere "$1$" and below diagonal we have $d-1$
subdiagonals completely filled by "$1$"; in all other places we have "$0$"
(the case with $d=2$ is shown in Eq.(\ref{matrix})).

The number of all distinct normally ordered {\bf sequences of words} of
length $s$ with allowed commutation relations is given by the following
partition function
\be \label{rs}
R_n(s,d)=\tilde{{\bf v}}_{\rm in}\left[\widehat{M}_n(d)\right]^{s-1}
{\bf v}_{\rm out}
\ee
where
\be \label{vec}
\tilde{{\bf v}}_{\rm in}=(\;\overbrace{1\; 1\; \ldots\; 1}^{n}\;)
\qquad \mbox{and}\qquad {\bf v}_{\rm out}=
\left.\left(
{\renewcommand{\arraystretch}{0.8} \begin{array}{c} 1 \\ 1 \\ \vdots
\\ 1 \end{array}} \right)\right\}n
\ee
For $s=1$ we have $R_n(1,d)=\tilde{{\bf v}}_{\rm in} {\bf v}_{\rm out}=2n$
as it should be.

The remaining sum in Eq.(\ref{2:trace}) is independent on $R_n(s,d)$, so
its calculation is very simple (see also Appendix A for some
generalizations):
\be \label{2:perm}
\mathop{{\sum}'}_{\{m_1,\ldots,m_s\}} \Delta\left[\sum_{i=1}^s
|m_i|-\mu\right] = 2^s\; C_{\mu-1}^{s-1}
\ee

Substituting Eq.(\ref{2:perm}) and Eq.(\ref{rs}) into Eq.(\ref{2:trace})
we get
\be \label{fin}
\begin{array}{ccl}
V_n(\mu)\equiv V_n(\mu,d) & = & \disp \sum_{s=1}^{\mu} 2^s\;
C_{\mu-1}^{s-1}\; R_n(s,d) \medskip \\
& = & \disp 2\tilde{{\bf v}}_{\rm in} (2\widehat{M}_n(d)+
\widehat{I})^{\mu-1}{\bf v}_{\rm out}\equiv 2\; {\rm Trace}\;
(2\widehat{M}_n(d)+\widehat{I})^{\mu-1}
\end{array}
\ee
where $\widehat{I}$ is the identity matrix.

Such a quantity is rather difficult to evaluate exactly. A reasonable
approximation is to replace (\ref{fin}) by
\be \label{fin1}
V_n^*(\mu)=2\sum_{i=1}^n (2\lambda_i+1)^{\mu-1}
\ee
where $\lambda_i$ are the eigenvalues of the matrix $\tilde{T}_n$ which can be
shown {\bf all to be real} (see the later discussion). In order to check the
validity of approximation (\ref{fin1}) we have considered the case
$\{n=3,\; d=2\}$
where the exact value reads
$$
V_3(\mu)=\left(\frac{15+7\sqrt{5}}{5}\right) (2-\sqrt{5})^{\mu-1} +
\left(\frac{15-7\sqrt{5}}{5}\right) (2+\sqrt{5})^{\mu-1}
$$
whereas the approximation (\ref{fin1}) gives
$$
V_3^*(\mu)=2(2-\sqrt{5})^{\mu-1}+2(2+\sqrt{5})^{\mu-1}+2(-1)^{\mu-1}
$$
It can be seen that this approximation works reasonably well even for small
values of $\mu$.

The value $V_n(\mu,d)$ is growing exponentially fast with $\mu$ and the
"speed" of this growth is clearly represented by the fraction
\be \label{fraction}
q(d)=\frac{V_n(\mu+1,d)}{V_n(\mu,d)}\bigg|_{\mu\gg 1}
\ee
which has the meaning of the effective coordinational number of the graph
$C({\cal LF}_n)$.

In the next section we present calculations of the asymptotic expression of
(\ref{fin1}) when $n\gg 1$.

\subsection{Calculation of Eigenvalues of Matrix $\widehat{M}_n(2)$}

Consider the determinant
\be \label{co:1}
a_n(\lambda)=\det\left(\widehat{M}_n - \lambda\widehat{I}\right)=
\left(
\begin{array}{rrrr}
-\lambda & 1        & 1        & \ldots  \\
1        & -\lambda & 1        & \ldots  \\
0        & 1        & -\lambda & \ldots  \\
\vdots   & \vdots   & \vdots   & \ddots  \\
\end{array}\right)
\ee
It satisfies the recursion relation
\be \label{co:2}
a_n(\lambda)=-(\lambda+1)a_{n-1}(\lambda)-(\lambda+1)a_{n-2}(\lambda)
\ee
with the boundary conditions
\be \label{co:3}
\left\{\begin{array}{ll}
a_0(\lambda)=1 \\ a_1(\lambda)=-\lambda
\end{array}\right.
\ee
For $\lambda>-1$ one may set
\be \label{co:4}
a_n(\lambda)=(\lambda+1)^{\frac{n-1}{2}}(-1)^n\varphi_n(\lambda)
\ee
which gives
\be \label{co:5}
\varphi_n(\lambda)=\sqrt{\lambda+1}\varphi_{n-1}(\lambda)-
\varphi_{n-2}(\lambda)
\ee

The general solution of (\ref{co:5}) satisfying the previously defined
boundary conditions (\ref{co:3}) is given in terms of Chebyshev's
polynomials of second kind
\be \label{co:6}
\varphi_n(\lambda)={\cal U}_{n+1}(\cos\vartheta)
\ee
where
\be \label{co:7}
\cos\vartheta=\frac{\sqrt{\lambda+1}}{2} \qquad
\left(0<\vartheta<\frac{\pi}{2}\right)
\ee
Therefore
\be \label{co:8}
\begin{array}{lll}
a_n(\lambda) & = & \disp (-1)^n(\lambda+1)^{\frac{n-1}{2}}\,
{\cal U}_{n+1}(\cos\vartheta) \medskip \\
& = & \disp (-1)^n(\lambda+1)^{\frac{n-1}{2}}\,
\frac{\sin(n+2)\vartheta}{\sin\vartheta}
\end{array}
\ee

The last expression enables us to obtain all the eigenvalues of the matrix
$\widehat{M}_n$. In fact, it is convenient to distinguish them according to 
the parity in $n$:

\begin{enumerate}
\item $n=2m+1$
\be \label{co:9}
\lambda_0=(-1)\quad (\mbox{$m$ such values});\qquad \lambda_k
=4\cos^2\frac{k\pi}{2m+3}-1\quad \Bigl(k=[1,m+1]\Bigr)
\ee
\item $n=2m$
\be \label{co:10}
\lambda_0=(-1)\quad (\mbox{$m$ such values});\qquad \lambda_k
=4\cos^2\frac{k\pi}{2m+2}-1\quad \Bigl(k=[1,m]\Bigr)
\ee
\end{enumerate}

Since in each case we have exactly $n$ states, this exhaust the complete
set of eigenvalues, showing they {\bf all are real} in the interval $[-1,3]$.
One also recovers the result obtained earlier in \cite{nevegr,jean} for the
asymptotics of the highest eigenvalue of matrix $\widehat{M}_n$ (in the limit
$n\gg 1$):
\be \label{co:11}
\lambda_{\rm max}= 4 \cos^2\frac{\pi}{n+2}-1\Bigg|_{n\gg 1}\approx
3-\frac{4\pi^2}{n^2}; \qquad (k=1)
\ee

Now we are in position  to compute the number of nonequivalent words
$V_n^{*}(\mu)$ of the primitive length $\mu$ in the locally free group
${\cal LF}_n(2)$ for $n\gg 1,\; n={\rm const}$ (see Eq.(\ref{fin1})).
Using the definition (\ref{fin1}) and Eqs.(\ref{co:9})--(\ref{co:10}) we
get for $n=2m+1$:
\be \label{ne:1}
V_n^*(\mu)=(n-1)(-1)^{\mu-1}+ 2\sum_{k=1}^{\frac{n+1}{2}}
\left(8 \cos^2\frac{k\pi}{n+2}-1 \right)^{\mu-1}
\ee
Define $\varphi_n(\mu)$ as follows:
$$
V_n^*(\mu)=(n-1)(-1)^{\mu-1}+2 \varphi_n(\mu)
$$

By means of Euler--Mac Laurin formula we may compute the asymptotic
expression of $\varphi_n(\mu)$:
\be \label{co:27}
\begin{array}{rcl}
\varphi_n(\mu)&=& \disp \frac{1}{2}\sum_{k=0}^{n+1}
\left(8\cos^2\frac{k\pi}{n+2}-1\right)^{\mu-1} \medskip \\
&=&\disp \frac{n+2}{2\pi}\int_0^{\pi}(8\cos^2x-1)^{\mu-1}dx-
\frac{1}{2}7^{\mu-1}
\end{array}
\ee
For for $\mu\gg 1$, $n$=const$\gg 1$ the last integral is evaluated by a
saddle point approximation which yields
\be
\varphi_n(\mu)=\frac{n+2}{2\pi}\frac{1}{2} \sqrt{\frac{\pi}{2\mu}}
7^{\mu-1}-\frac{1}{2}7^{\mu-1}
\ee
Thus, for the number of nonequivalent words in the locally free group
${\cal LF}_n(2)$ we have the following limiting behavior:
\be \label{co:28}
\disp \lim_{\scri \shortstack{\\
$\mu\to\infty$ \\ $n={\rm const}\gg 1$}}
\frac{1}{\mu}\ln V_n^*(\mu)= \ln7
\ee
and Eq.(\ref{fraction}) gives
$$
q(d=2)= 7
$$

Hence, the graph corresponding to the locally free group can be viewed as an
effective tree with the branching number $q=7$.

\section{Approximate Statistical Approach for Words Enumeration in Braid
Group $B_n$}

The construction of an effective algorithm for enumeration of the words in
the braid group $B_n$ for $n>2$ is of the most intriguing problems in
group theory.

In the present section we propose an {\it approximate statistical
approach} for enumeration of all distinct primitive words in the group
$B_n$ for $n\gg 1$ which exploits some properties of locally free groups
${\cal LF}_n$ considered above.

The main idea is as follows. Let us deal with the sequences of words in
the braid group $B_n$ from the point of view of the locally free group
${\cal LF}_n^{\rm err}$ "with errors". To be more specific let us start
with the following example:

\noindent {\bf Example 2.} Write a random word $W$ in the group $B_7$
consisting of 8 letters. Let this word be for instance:
$$
W\,=\,(\sigma_1)^{-1}\, \sigma_4\, (\sigma_5)^{-1}\, (\sigma_6)^{-1}\,
\sigma_5\, \sigma_1\, \sigma_6 (\sigma_2)^{-1}
$$
We reduce this word to the primitive one in two steps.

1. On the first step we act in the same way as in the case of locally free
group ${\cal LF}_7$ and push all generators with smaller indices to the left
{\it assuming that nearest neighbors do not commute at all}. We get:
$$
W_{\rm reduced}\, =\, (\sigma_2)^{-1}\, \sigma_4\,
(\sigma_5)^{-1}\, \underbrace{(\sigma_6)^{-1}\, \sigma_5\, \sigma_6}_
{\mbox{$\sigma_5\, \sigma_6\, (\sigma_5)^{-1}$}}
$$

2. Now we can apply the Yang-Baxter relations to the triple
$(\sigma_6)^{-1}\, \sigma_5\, \sigma_6$ and obtain after the cancellation
of $(\sigma_5)^{-1}$ and $\sigma_5$ the primitive word
$$
W_p\,=\,(\sigma_2)^{-1}\, \sigma_4\,\sigma_6\, (\sigma_5)^{-1}
$$

The first step of the braid contracting procedure completely coincides with
what we did for the locally free group, while the second step we could
regard (approximately, of course) as follows:

{\it Consider some pair, for instance, $(\sigma_6)^{-1}\, \sigma_5$. We
commute it with the probability $p$. Such commutation we denote as an {\bf
error}. The probability to meet the generator $\sigma_i$ in the Markov
chain with uniform distribution over the generators in the braid group
$B_n$ is of order of $p=\frac{1}{2n}$. Later on we consider more general
case taking $p$ as the variational parameter.}

So, let $V_n^{\rm braid}(\mu)$ be the number of all primitive words of length
$\mu$ in the braid group $B_n$. Our main idea is as follows: we would like to
relate the quantity $V_n^{\rm braid}(\mu)$ to the number of primitive words
in the "group" ${\cal LF}_n^{\rm err}(2)$ averaged over the uniform
distribution of "errors" in commutation relations.

It should be pointed out that the object ${\cal LF}_n^{\rm err}(2)$ cannot be
considered as a real group anymore. So everywhere below we understand under
${\cal LF}_n^{\rm err}(2)$ just the ensemble of random words written in
terms of the alphabet with specific commutation relations among the letters.

\subsection{Statistics of Words with "Errors" in Locally Free Groups}

The methods of theoretical description of the systems with disorder are
rather well developed, especially in regard to the investigation of {\it
spin glass models} \cite{mezard}.

Central for these methods is the concept of {\it self-averaging} which can
be explained as follows. Take some additive function $F$ (the free energy,
for instance) of some disordered system. The function $F$ is the
self-averaging quantity if the observed value, $F_{obs}$, of any
macroscopic sample of the system coincides with the value $F_{av}$ averaged
over the ensemble of disorder realizations:
$$
F_{obs} = \left<F\right>_{av}
$$
The phenomenon of self-averaging takes place in the systems with
sufficiently weak long-range correlations: in this case only $F$ can be
considered as a sum of contributions from different volume domains,
containing statistically independent realizations of disorder (for more
details see \cite{lifsh}).

The central technical problem of systems with {\it quenched} disorder deals
with the calculation of the free energy $F(\mu)$  averaged over the randomly
distributed quenched pattern. In our case we could associate the number of
topologically different words with the partition function, hence the free
energy would be $F(\mu)=-\left<\ln V_n^{\rm err}(\mu)\right>$ and the
"quenched pattern" is just the set of "errors" in commutation relations.

A problem closely related to that mentioned above arises when averaging the
correlation functions of some statistical system over the disorder. In this
case the computations are based on finding the averaged density of states
of some random matrix over the prescribed distribution of random entries.
Below we show that the calculation of the mean value $\left<V_n^{\rm
err}(\mu)\right>$ belongs precisely to this class of problems.
\medskip

\begin{conjecture} The number of nonequivalent primitive words,
$V_n^{\rm braid}(\mu)$, of length $\mu$ in the braid group $B_n$ can be
estimated in the limits $n={\rm const}\gg 1,\; \mu\gg 1$ as follows:
\be \label{esim}
V_n^{\rm braid}(\mu)\approx \left<V_n(\mu,d=2,p)\right>
\ee
where $V_n(\mu,d=2,p)$ is the number of all distinct primitive words of
length $\mu$ with the "{\bf errors}" in the commutation relations in the
locally free group ${\cal LF}_n(2)$. We allow to commute the neighboring
generators with the probability $p$ and the averaging is performed over the
uniform probability distribution of "errors". {\rm The question
concerning the choice of $p$ is considered below.}
\end{conjecture}

In support of our conjecture we bring the numerical computations performed
in the work \cite{jean}, where we have constructed the (right-hand) random
walk (the random word) on the group ${\cal G}_n=\left\{{\cal LF}_n^{\rm
err},\, B_n\right\}$ with a uniform distribution over generators
$\{g_1,\ldots, g_{n-1}, g_1^{-1},\ldots,g_{n-1}^{-1}\}\in {\cal G}_n$. It
means that with the probability $\frac{1}{2n-2}$ we have added the element
$g_{\alpha_N}$ or $g_{\alpha_N}^{-1}$ to the given word of $N-1$ generators
(letters) from the right-hand side. In \cite{jean} the following question
has been raised:  what is the averaged length of the primitive path
$\left<\mu\right>$ for the $N$--step random walk on the group ${\cal G}_n$?

In the Table 1 we show the results of numerical simulations carried on in
\cite{jean} for the expectation value $\left<\mu\right>/N$ of the $N$--step
random walk on the "locally free structure with errors", ${\cal LF}_n^{\rm
err}(2)$, and compare them to the same value for the $N$--step random walk
on the braid group $B_n$.

{\small
\begin{center}
{\bf Table 1.}
\nopagebreak

\medskip
\begin{tabular}{|c||c||c|} \hline
& \multicolumn{1}{|c||}{${\cal LF}_n^{\rm err}$ with
$p=1/5$}
& \multicolumn{1}{|c|}{$B_n$} \\ \hline\hline
$n$ & $\left<\mu\right>/N$ & $\left<\mu\right>/N$
\\ \hline
5 & 0.55 & 0.49
\\ \hline
10 & 0.58 & 0.56
\\ \hline
20 & 0.59 & 0.59
\\ \hline
50 & 0.60 & 0.61
\\ \hline
100 & 0.60 & 0.61
\\ \hline
200 & 0.61 & 0.61
\\ \hline
\end{tabular}
\end{center}
}

We have found asymptotically very good correspondence of the mean values
$\left<\mu\right>/N$ for the braid group and the "locally free structure with
the errors" for $p=\frac{1}{5}$.

Of course, our conjecture is not exact and has a "mean--field" nature
because we admit two subsequent generators $(\sigma_{\alpha_k},\,
\sigma_{\alpha_{k+1}})$ with nearest neighbor indices $\alpha_{k+1}=
\alpha_k \pm 1$ to commute with the probability $p$ regardless the value of
the generator $\sigma_{\alpha_{k+2}}$ in the sequence of letters in the word.
Hence the problem of taking right value of $p$ appears.

Two limiting cases are trivial:

(i) For $p=1$ we have a complete commutative group and the obvious
inequality is fulfilled
\be \label{esim1}
V_n^{\rm braid}(\mu) \ge \left<V_n(\mu,d=2,p=1)\right>
\ee

(ii) For $p=0$ we return to locally free group ${\cal LF}_2$, for which we
should have
\be
V_n^{\rm braid}(\mu) \le V_n(\mu,d=2,p=0)
\ee

In the framework of the mean-field approximation and taking into account
Eq.(\ref{esim}) we claim:
\be
\left<V_n(\mu,d=2,p)\right>\Big|_{p\to 1^-}\le V_n^{\rm braid}(\mu) \le
V_n(\mu,d=2,p=0)
\ee

We show below that in the limit $n\to\infty, \mu\to\infty$ the following
equality takes place
\be
\left<V_n(\mu,d=2,p)\right>\Big|_{p\to 1^-}= V_n(\mu,d=2,p=0)
\ee

\noindent {\bf The Main Statement.} {\it For the value $V_n^{\rm
braid}(\mu)$ we have the following asymptotic behavior}
\be \label{conj}
\lim_{\scri \shortstack{\\$\mu\to\infty$ \\ $n\to\infty$}}
\frac{1}{\mu} \ln V_n^{\rm braid}(\mu) =
\left[\frac{1}{\mu} \ln \left<V_n(\mu,d=2,p)\right>\right]_{\scri
\shortstack[l]{independent \\ on $p$}}=\ln 7
\ee
\medskip

It is easy to understand that the number of nonequivalent primitive
words $V_n(\mu,d=2,p)$ in the "locally free group with errors" can be
calculated by means of averaging of Eq.(\ref{fin}) if we slightly change
the matrix $\widehat{M}_n$ replacing it by the random incidence matrix
$\widehat{M}_n$:

\be \label{co:12}
\widehat{M}_n^{\rm err}(d=2)=
{
\renewcommand{\arraystretch}{0}
\begin{tabular}{|c||c|c|c|c|c|c|c|} \hline
\strut \rule{0pt}{15pt} & $f_1$ & $f_2$ & $f_3$ & $f_4$ & $\ldots$ &
$f_{n-1}$ & $f_n$ \\ \hline \rule{0pt}{2pt} & & & & & & & \\ \hline
\strut \rule{0pt}{15pt} $f_1$ & 0 & 1 & 1 & 1 & $\ldots$ & 1 & 1 \\
\hline
\strut \rule{0pt}{15pt} $f_2$ & $x_{n-1}$ & 0 & 1 & 1 & $\ldots$ & 1 & 1
\\ \hline
\strut \rule{0pt}{15pt} $f_3$ & 0 & $x_{n-2}$ & 0 & 1 & $\ldots$ & 1 & 1
\\ \hline
\strut \rule{0pt}{15pt} $f_4$ & 0 & 0 & $x_{n-3}$ & 0 & $\ldots$ & 1 & 1
\\ \hline
\strut $\vdots$ & $\vdots$ & $\vdots$ & $\vdots$ & $\ddots$ &
$\ddots$ & $\vdots$ & $\vdots$ \\ \hline
\strut \rule{0pt}{15pt} $f_{n-1}$ & 0 & 0 & 0 & 0 & $\ddots$ & 0 & 1 \\
\hline
\strut \rule{0pt}{15pt} $f_n$ & 0 & 0 & 0 & 0 &  & $x_1$ & 0 \\
\hline \end{tabular}
}
\ee
\medskip

\noindent It is now parametrized by the random sequence of "0" or
"1", i.e.
\be \label{co:13}
\left\{x^{(n)}\right\}=\left\{x_{n-1},x_{n-2},\ldots,x_2,x_1\right\}
\ee
where
\be \label{co:14}
\begin{array}{l} \mbox{Prob}(x_j=1)=1-p \\
\mbox{Prob}(x_j=0)=p \end{array}
\ee

\subsection{Density of States of Random Operator and Averaged Number of
Nonequivalent Words}

The determinant of the matrix $\widehat{M}_n^{\rm err}(d=2)-\lambda
\widehat{I}$ (see Eq.(\ref{co:12})) satisfies the random recursion
relation
\be \label{co:30}
a_{k+1}+a_k(\lambda+x_k)+a_{k-1}(1+\lambda)x_k=0
\ee
Introducing the Ricatti--like variable
$$
\rho_k=\frac{a_{k+1}}{a_k}
$$
we arrive at the recursion relation
\be \label{co:31}
\rho_k=-(\lambda+x_k)-\frac{(1+\lambda)x_k}{\rho_{k-1}} \qquad k\in [0,n]
\ee
with the boundary condition $\rho_0=-\lambda$.

For the continuous sequence of $\{1\}$, i.e. $\{x^{(n)}\}=\{1\;1\;1\;1\;
\ldots\;1\}$ we have the following (nonrandom) transformation
\be \label{co:32}
\rho_{k+1}=-(\lambda+1)\left(1+\frac{1}{\rho_k}\right)
\ee
As soon as a zero appears in the random sequence, $\rho_k$ in (\ref{co:31})
is set to $-\lambda$ which coincides precisely with the initial value
$\rho_0$. Since one returns back to the initial value, the process can be
easily iterated for arbitrary random sequences $\{x^{(n)}\}$ when $n\gg 1$.
Such a property of the map (\ref{co:31}) is equivalent to the factorization
of the determinant $a_n(\lambda)$ of the random matrix $\widehat{M}_n^{\rm
err}(d=2)-\lambda\widehat{I}$:

\noindent {\bf Example 2}. Consider the sequence
\be \label{co:15}
\left\{x^{(n)}\right\} = (\underbrace{1\;1\;\ldots\;1}_{l_1}\;
\underbrace{0\;0\;\ldots\;0}_{m_1} \;\underbrace{1\;\ldots\;1}_{l_2})
\ee
The corresponding determinant $a_n(\lambda)$ factorizes:
\be \label{co:16}
a_n(\lambda)=\prod_{\{l_j\}}a_{l_j}(\lambda)
\prod_{\{m_j\}}(-\lambda)^{m_j-1}
\ee

It is worth pointing out that a recursion relation similar to (\ref{co:32})
appears also in the study of the binary product of random $2\times 2$
matrices where one of the matrix is singular \cite{lima}. Such a structure
also occurs in the case of Ising model in a random magnetic field---see
\cite{derrida}. Following Derrida and Hilhorst \cite{derrida}, one can
write down the invariant measure associated to (\ref{co:31}):
\be \label{co:33}
{\cal P}(\rho)=\sum_{k=1}^{\infty}p(1-p)^k \delta(\rho-\rho_k)+
(1-p)\sum_{k=1}^{\infty}p^k \delta(\rho-\rho_0)
\ee
The first (the second) term comes from the complete sequences of $\{1\}$
($\{0\}$) of arbitrary length $k=1,2,\ldots$. From the invariant measure
one can compute $\overline{\ln a(\lambda)}$ which can be interpreted either
as a Lyapunov exponent or as the free energy (depending on the physical
context). So, we get:
\be \label{co:34}
\begin{array}{lll}
\overline{\ln a(\lambda)}\equiv \disp \lim_{n\to\infty} \overline{\ln
a_n(\lambda)} & = & \disp \int {\cal P}(\rho) \ln\rho d\rho \\ & = & \disp
\sum_{k=1}^{\infty}p^2(1-p)^{k-1}\ln a_k(\lambda)
\end{array}
\ee

Returning to our problem, we may use this expression to write down the
averaged density of states
\be \label{co:35}
\overline{\rho (\lambda)}=\frac{1}{\pi}\frac{\partial}{\partial \lambda}
{\rm Im}\,\overline{\ln a(\lambda)}\equiv \sum_{k=1}^{\infty}p^2(1-p)^{k-1}
\rho_k(\lambda)
\ee
where $\disp\rho_n(\lambda)\equiv \frac{1}{\pi}{\rm Im}\,\ln a_n(\lambda)$ is
the density of states of a pure system (i.e. without randomness) of length
$n$. From the density of states we can find the average number of words
in the limit $n\to \infty$

Define $\left<V^*(\mu)\right>$ as follows:
\be \label{co:24}
\left<V^*(\mu)\right>=\lim_{n\to\infty}
\frac{\left<V_n^*(\mu)\right>}{n}=2\int (2\lambda+1)^{\mu-1}
\overline{\rho(\lambda)}\,d\lambda
\ee
where the integration over $\lambda$ runs over the whole spectrum. Let us
repeat once more that the function $\overline{\rho(\lambda)}$ is the
density of states of the random matrix $\widehat{M}_n(d=2)-\lambda
\widehat{I}$ ($n\to \infty$ averaged over the disordered pattern
$\{x^{(n)}\}$.

The content of Eq.(\ref{co:35}) is as follows. The density of states
$\overline{\rho(\lambda)}$ can be obtained by averaging the spectrum of
the regular case "weighted" with associated sequences of $\{1\}$:
\be \label{co:17}
\mbox{Prob}\left\{x=(0\;\underbrace{1\;1\;1\;1\;\ldots\;1\;1\;
1}_{\mbox{\scriptsize complete set of $\{1\}$}} \;0)\right\}=p^2(1-p)^n
\ee

One should also add the contribution coming from zero's energy state
corresponding to sequences of $\{0\}$.
The resulting expression reads:
\be \label{co:18}
\begin{array}{lll}
\disp\overline{\rho(\lambda)} & = & \disp \sum_{m=1}^{\infty}
p^2(1-p)^{2m-1} \left\{m\delta(\lambda+1)+\sum_{k=1}^{m}
\delta\left(\lambda+1-4\cos^2\frac{k\pi}{2m+2}\right)\right\} \medskip \\
& + & \disp \disp\sum_{m=0}^{\infty}p^2(1-p)^{2m}
\left\{m\delta(\lambda+1)+\sum_{k=1}^{m+1}
\delta\left(\lambda+1-4\cos^2\frac{k\pi}{2m+3}\right)\right\}
\end{array}
\ee
which may be rewritten as
\be \label{co:19}
\overline{\rho(\lambda)} = \frac{1-p}{2-p}\delta(\lambda+1)+
\sum_{n=0}^{\infty}p^2(1-p)^n\sum_{k=1}^{\left[\frac{n+2}{2}\right]}
\delta\left(\lambda+1-4\cos^2\frac{k\pi}{n+3}\right)
\ee
where $[x]$ denotes the integer part of $x$.

Using Eq.(\ref{co:19}) we may check that the function
$\overline{\rho(\lambda)}$ is properly normalized:
$$
\int\limits_{-\infty}^{+\infty} \overline{\rho(\lambda)} =
\int\limits_{-1}^{3} \overline{\rho(\lambda)} = 1
$$

Returning to (\ref{co:24}) we get
\be \label{co:25}
\left<V^*(\mu,p)\right>=2\left(\frac{1-p}{2-p}\right)(-1)^{\mu-1}+
2\sum_{n=0}^{\infty}p^2 (1-p)^n\, S_{n+2}(\mu)
\ee
where
\be \label{co:26}
S_n(\mu)=\sum_{k=1}^{\left[\frac{n}{2}\right]}
\left(8 \cos^2\frac{k\pi}{n+1}-1 \right)^{\mu-1}
\ee
(compare to (\ref{ne:1})).

In order to check the algebra we have computed $\left<V^*(\mu)\right>$
for small values of $\mu$. One gets:
$$
\left<V^*(1)\right>=\left<V^*(2)\right>=2
$$
which can be readily obtained through a direct calculation of $\disp
\lim_{n\to\infty} 2\mbox{Trace}\left(2\widehat{M}_n+\widehat{I}\right)^{\mu-1}$. We
are however mainly interested in the limit $\mu\to\infty$.

Using (\ref{co:27})--(\ref{co:28}) we may resum the series (\ref{co:25}) by
isolating the contribution $n<\mu$ and $n>\mu$. After some algebra one
obtains for $\mu<n+1$:
\be \label{co:28a}
S_n(\mu)=(n+1)\sum_{p=1}^{\mu-1}\,C_{\mu-1}^p\,(-1)^{\mu-p}
C_{2p-1}^{p-1} -
\frac{1}{2}\left\{7^{\mu-1}+(-1)^{\mu}\right\}-\left[\frac{n}{2}\right]
(-1)^{\mu}
\ee
which gives
\be \label{co:28b}
S_n(\mu)\Big|_{\mu\gg 1}\simeq
\left\{\begin{array}{cc}
7^{\mu} & n_0<n<\mu \medskip \\ (n+1)\, 7^{\mu} & \mu<n
\end{array}\right.
\ee
where $n_0$ is some constant of order of unity. The Eq.(\ref{co:28a}) is
plotted in Fig.\ref{fig:2}a,b.

Thus, we can rewrite Eqs.(\ref{co:25})--(\ref{co:26}) as follows
\be \label{co:28c}
\left<V^*(\mu,p)\right>=2\left(\frac{1-p}{2-p}\right)(-1)^{\mu-1}+
\sum_0^{\mu}p^2(1-p)^n S_{n+2}(\mu)+
\sum_{\mu}^{\infty} p^2(1-p)^n S_{n+2}(\mu)
\ee
The corresponding behavior of the function $Q(p|\mu)$ where
\be \label{co:plot}
Q(p|\mu)=\frac{\ln \left<V^*(\mu,p)\right>}{\mu} \qquad (0<p<1)
\ee
is shown in Fig.\ref{fig:3} for few fixed values $\mu=\{10,\,30,\,150\}$.

The plot in Fig.\ref{fig:3} enables us to come to the following conclusion.
If the number of "errors" is small ($p\to 0^+$), the volume of the group
grows exponentially with the Lyapunov exponent $\ln7$ (for $\mu\to\infty$).
For the arbitrary number of "errors", $p$, the corresponding Lyapunov
exponent approaches the same value $\ln 7$ for all $p<1$ in the limit
$\mu\to\infty$ and exhibits a singular behavior just at the point $p=1$
(which corresponds to the completely commutative group).

The asymptotic expression (\ref{co:28c}) allows us to conclude that the
limit behavior of the function $V^*(\mu)$ is independent on $p$, $\forall
p\in ]0,1[$ and is the same as for the locally free group ${\cal LF}_n$
without any errors. This fact supports our conjecture (\ref{conj}).

It should be emphasized that these results are expected to hold only in the
thermodynamic limit $n\to \infty$. It would be more desirable to consider
the limit in which the number of generators $n$ is kept fixed and the
length of the word, $\mu$, is much larger than $n$.

\subsection{Functional Equation, Continued Fractions and Invariant Measure}

The behavior of the spectral density of our model is very similar to the
one encountered in the study of harmonic chains with binary random
distribution of masses. This problem, which goes back to Dyson has been
investigated by Domb {\it et al} \cite{domb} and then  thoroughly
discussed by Nieuwenhuizen and Luck \cite{huisen}. One considers a chain of
oscillators where the masses can take two values:
$$
\left\{\begin{array}{cl}
m & \mbox{with the probability $1-p$} \\
M>m & \mbox{with the probability $p$}
\end{array}\right.
$$

In the limit $M\to\infty$ the system breaks into islands, each of which
consisting of $n$ light masses surrounded by two infinite heavy masses. The
probability of occurence of such an island is $p^2(1-p)^n$. There is
clearly a mapping to our model if one replaces the sequences of heavy and
light masses by the sequences of "0" and "1". Many results may therefore be
borrowed from the works \cite{domb,huisen}. In particular, by adapting the
calculations of Nieuwenhuizen and Luck to our case one may rewrite the
integrated density of states
$$
\overline{{\cal N}(\lambda)}=\int_{-\infty}^{\lambda}
\overline{\rho(\lambda')}d\lambda'
$$
in the form
\be \label{co:20}
\overline{{\cal N}(\lambda)}=1-\frac{p}{(1-p)^2}\sum_{n=1}^{\infty}
(1-p)^{\rm Int\left(\frac{n\pi}{\vartheta}\right)}
\ee
where the relation between $\lambda$ and $\vartheta$ is given in
Eq.(\ref{co:7}).

For $\lambda\to -1$ one gets $\disp \overline{{\cal N}(\lambda)}\to
\frac{1-p}{2-p}$ which corresponds to the contribution of the states
$\lambda=-1$ at the bottom of the spectrum.

At the upper edge of the spectrum (namely for $\lambda\to 3^{-}$) one gets
$\overline{{\cal N}(\lambda)}\to 1$ which means that all the states are
counted. Equation (\ref{co:20}) shows that the behavior around
$\lambda=3$ (corresponding to $\vartheta=0$) is in fact dominated by the
first term ($n=1$) of the series. One has:
\be \label{co:21}
\begin{array}{lll} \overline{{\cal N}(\lambda)} & \simeq & \disp
1-\frac{p}{(1-p)^2}(1-p)^{\frac{2\pi}{\sqrt{3-\lambda}}} \medskip \\ &
\equiv & \disp 1-\frac{p}{(1-p)^2} \exp\left[\frac{2\pi}{\sqrt{3-\lambda}}
\ln(1-p)\right]
\end{array}
\ee

The behavior (\ref{co:21}) signals the appearence of Lifshits' singularity
in the density of states. A more precise analysis shows that this result is
in fact modulated by a periodic function \cite{huisen}.

Equation (\ref{co:20}) displays many interesting features. In particular,
the function $\overline{{\cal N}(\lambda)}$ occurs in the mathematical
literature as a generating function of the continued fraction expansion of
$\disp\frac{\pi}{\vartheta}$.

Let us briefly sketch this connection. Consider the continued fraction
expansion
\be \label{co:23}
\frac{\pi}{\vartheta}=\frac{1}{\disp c_0+
\frac{1}{\disp c_1+
\frac{1}{\disp c_2+\ldots}}}
\ee
where all $c_n$ are natural integers. Truncating this expansion at level
$n$ we get a rational number $\disp \frac{p_n}{q_n}$ which converges to
$\disp\frac{\pi}{\vartheta}$ when $n\to \infty$.

A theorem of B\"ohmer \cite{bohmer} states that the generating function of
the integer part of $\disp\frac{\pi}{\vartheta}$
$$
G(z)=\sum_{n=1}^{\infty}z^{\rm Int\left(\frac{n\pi}{\vartheta}\right)}
$$
is given by the continued fraction expansion
$$
G(z)=\frac{z}{1-z} \frac{1}{\disp A_0+ \frac{1}{\disp A_1+
\frac{1}{\disp A_2+\ldots}}}
$$
where
$$
A_n(z)=\frac{\left(\disp\frac{1}{z}\right)^{q_n}-
\left(\disp\frac{1}{z}\right)^{q_{n-2}}}
{\left(\disp\frac{1}{z}\right)^{q_{n-1}}-1}
$$
and $q_n$ is the denominator of the fraction $\disp \frac{p_n}{q_n}$
approximating the value $\disp\frac{\pi}{\vartheta}$.

In order to make connection with our problem it is sufficient to set
$z=1-p$ and express $\overline{{\cal N}(\lambda)}$ in terms of $G(z)$.

Equation (\ref{co:20}) shows that the invariant measure is a very singular
object. However it satisfies a simple functional equation reminiscent of
that which arises in the theory of automorphic forms. The equation can be
derived either by a Dyson-Schmidt approach (see, for instance \cite{lifsh})
or just by looking at the explicit expression of ${\cal P}(\rho)$.

It is in fact simpler to work with the rescaled variable
$$
z_n=-\frac{1}{\sqrt{\lambda+1}}\rho_n
$$
which satisfies the recursion relation equivalent to Eq.(\ref{co:32})
\be \label{co:36}
z_n=\mu-\frac{1}{z_{n-1}}
\ee
where $\mu=\sqrt{\lambda+1}$. The transformation (\ref{co:36}) may be
obtained from the two matrices belonging to the group $SL(2,{\R})$
$$
\widehat{T}=\left(\begin{array}{cc} 1 & \mu \\ 0 & 1 \end{array} \right);
\qquad
\widehat{S}=\left(\begin{array}{cc} 0 & -1 \\ 1 & 0 \end{array}  \right),
$$

The $SL(2,{\R})$--transformation $\widehat{T}\widehat{S}= \left(a\; b \atop c\;
d\right)$ acts on $z$ by the fractional linear transformation
\be \label{co:37}
z_n=\frac{a z_{n-1}+b}{c z_{n-1}+d}=\mu-\frac{1}{z_{n-1}}.
\ee
The invariant measure of (\ref{co:36}), which may be rewritten as
\be \label{co:38}
{\cal P}(z)=\sum_{n=0}^{\infty}p(1-p)^n\delta\left(z-[\widehat{S}\widehat{T}]^n
z_0\right)
\ee
can easily be shown to satisfy the fundamental equation
\be \label{co:39}
{\cal P}(z)=\frac{1}{1-p}\;(cz+d)^2\;{\cal P}(\widehat{S}\widehat{T}z)
\ee
up to some singular terms.

By suitable rescaling it is in fact possible to absorb the prefactor
$1/(1-p)$ and rewrite Eq.(\ref{co:39}) as
\be \label{co:40}
{\cal P}(\widehat{g}z)=(cz+d)^{-2}\,{\cal P}(z)
\ee
where $\widehat{g}\in SL(2,{\R})$.

An analytical continuation of this expression into the complex $z$--plane
would eventually permit one to interpret ${\cal P}$ as an automorphic form.
>From the theory of automorphic functions it is well known that
Eq.(\ref{co:40}) is satisfied by the Dedekind modular function $\eta(z)$:
$$
\eta(z)=\frac{1}{(cz+d)^2}\,\eta\left(\frac{az+b}{cz+d}\right) \qquad
(ad-bc=1)
$$

Such objects although perfectly smooth in the upper half--plane ${\rm Im} z>0$
display highly non--trivial fractal behavior on the boundary ${\rm Im} z=0$
(see, for instance \cite{bog,berry}).

Another interesting connection which would be worth investigating is the
fact that $\widehat{S}$ and $\widehat{T}$ generate the so-called Hecke group
$\Gamma(h)$ for $h=2\cos\frac{\pi}{q}$ ($q\ge 3$ is integer).
Surprisingly, these values of $h$ coincide with a subset of the spectrum
of the matrix $\widehat{M}_n$ (see Eq.(\ref{co:9})).
\medskip

\section{Final Remarks}
\subsection{The Geometrical View on the Word Enumeration  Problem}

The number of primitive words in the locally free or braid groups allows a
rather straightforward geometrical description. Namely, the matrix
$\widehat{M}_n(d=2)$ can be regarded as the transfer matrix for the model
of a "biased Levy--flight"--like ("BLF"--like) one--dimensional random
walk on the finite support. Actually, let us compute the statistical
sum of the process described below. Take $n$ integers on the line:
$1,2,\ldots,n$ and consider the random walk when the walker can jump with
equal probabilities from the vertex with the coordinate $m_1$ ($1\le
m_1<n$) to:

(i) each vertex with the coordinate $m_2$ ($m_2\in[m_1+1,\, n]$);

(ii) the vertex with the coordinate $m_2=m_1-1$ (i.e. one step back).

The corresponding process is represented schematically in Fig.\ref{fig:4}.
\medskip

Analogously, we can associate the random operator $\widehat{M}_n^{\rm err}$
with the transfer matrix of the generalized BLF--like random process
which is described via the same rules (i) and (ii) but with additional
requirement that the jump (ii) is blocked with probability $p$ and
allowed with probability $1-p$, independently of the position of the vertex.

\subsection{Conclusion}
We have proposed a statistical method for enumerating the primitive words in
the braid group $B_n$ based on the consideration of {\it locally free groups
with errors in commutation relations}. We brought arguments in support
of the conjecture that the number of long primitive words in the braid
group is not sensitive to the precise local commutation relations. We
discussed the connection of the abovementioned problems with the
conventional random operator theory, localization phenomena and statistics
of systems with quenched disorder, and showed the connection between some
particular problems of random matrix theory and the theory of automorphic
functions.

We believe that the problem of discovering the integrable models associated
with the proposed locally free groups and developing the corresponding
conformal field theory could help establish a bridge between the
statistics of random walks on the noncommutative groups, spectral theory on
multiconnected Riemann surfaces, and topological field theory.

\subsubsection*{Acknowledgments}

We are very grateful to J. Desbois for elucidating for us many questions
concerning the limiting behavior of random walks on locally free and braid
groups; we would like to thank as well M. Tsypin for useful remarks and
for help in the numerical confirmation of some of our conjectures. S.N.
acknowledges the fruitful discussions with S. Fomin, L. Pastur, Ya.
Sinai and A. Vershik on many aspects of the work. We highly appreciate the
assistance of O.Martin in the final preparation of the paper and critical
reading of the manuscript.

\newpage

\section*{Appendix A}

The calculation of the number of distinct primitive words $V_n^{\rm
p}(\mu)$ can be easily extended to the case of the group ${\cal LF}_n^{\rm
perm}$ which is defined by the old relations:
\begin{itemize}
\item[(a)] Each pair $(g_j, g_k)$ generates the free subgroup of the group
${\cal LF}_n^{\rm perm}$ if $|j-k|<2$;
\item[(b)] $g_j g_k=g_k g_j$ for  $|j-k|\ge 2$
\end{itemize}
completed with the additionnal requirement\footnote{Compare to the definition of
the group of permutations.} $(g_i)^p=1$ which may be rewritten also
as:
$$
g_j=1 \qquad |\mbox{mod $p$}| \quad \forall j\in [1,n]
\eqno (A.1)
$$

The calculation of the number of distinct primitive words, $V_n^{\rm
p}(\mu)$, can be carried out using Eq.(\ref{2:trace}):
$$
V_n^{\rm perm}(\mu)=\sum_{s=1}^{\mu}
R_n(s,d)\mathop{{\sum}''}_{\{m_1,\ldots,m_s\}} \Delta\left[\sum_{i=1}^s
|m_i|-\mu\right]
\eqno (A.2)
$$
where "double prime" means that the sum runs over all $m_i\neq 0$
consistent with Eq.(A.1).

Taking into account that the summation over all $\{m_1,\ldots,m_s\}$
obeying the condition (A.1) is independent on computation of $R_n(s)$, we
have to replace the Eq.(\ref{2:perm}) by the following one
$$
\begin{array}{rcl}
\disp {\cal N}\equiv\mathop{{\sum}''}_{\scri \shortstack{\\
$\{m_1,\ldots,m_s\}$ \\ $|\mbox{mod $p$}|$}}
\Delta\left[\sum_{i=1}^s |m_i|-\mu\right] & = & \disp
\frac{2^s}{2\pi i}\oint\limits_{(C)} dz\;
z^{-1-\mu}\left(\sum_{m=1}^{p-1}z^m\right)^s
\medskip \\ & = & \disp
\frac{2^s}{2\pi i}\oint\limits_{(C)} dz\;
z^{-1-\mu+s}\left(\frac{1-z^{p-1}}{1-z}\right)^s
\end{array}
\eqno (A.3)
$$
where the contour $C$ encloses the origin of the complex plane $z$

We used the integral representation of Kronecker $\Delta$--function
$$
\Delta(x)=\frac{1}{2\pi i}\oint\frac{dz}{z^{1+x}}=
\left\{\begin{array}{ll} 0 & x\neq 0 \\ 1 & x=0
\end{array}\right.
\eqno (A.4)
$$

Let us consider three special cases:

\begin{enumerate}
\item $p\to\infty$. From Eq.(A.3) we get
$$
{\cal N}=\frac{2^s}{2\pi i}\oint dz\; z^{-1-\mu+s}(1-z)^{-s}=
2^s\frac{(\mu-1)!}{(s-1)!(\mu-s)!}
\eqno (A.5)
$$
which coincides with Eq.(\ref{2:perm}).

\item $p=2$. From Eq.(A.3) we get
$$
{\cal N}=\frac{2^s}{2\pi i}\oint dz\; z^{-1+\mu-s}=2^s \,\delta_{\mu,s}
\eqno (A.6)
$$

\item $p=3$. From Eq.(A.3) we get
$$
{\cal N}=\frac{2^s}{2\pi i}\oint dz\; z^{-1+\mu-s}(1+z)^s=
2^s\frac{s!}{(2s-\mu)!(\mu-s)!}
\eqno (A.7)
$$
\end{enumerate}

Substituting Eqs.(A.5)--(A.6) and Eq.(\ref{rs}) into Eq.(\ref{2:trace}) we
may easily compute the corresponding values $V_n^{\rm perm}(\mu)$.

\newpage

\section*{Appendix B}

The computation of the averaged density of states $\overline{\rho
(\lambda)}$ of the random matrix $\widehat{M}^{\rm err}_n-\lambda
\widehat{I}$ belongs to a class of problems  with the "quenched
disorder" (see the discussion in the section 4.1). Another class of
problems closely related to the mentioned one arises when averaging the
partition function (but not the free energy) over the disorder. Problems
corresponding to the case of {\it annealed} disorder usually seem to be
simpler from computational point of view than that of {\it quenched}
disorder but usually the thermodynamic behavior of systems with annealed
disorder is less rich and less interesting than that of systems of quenched
disorder. However in our case it would be very desirable to compute the
value $\overline{a_n(\lambda)}$ where $a_n(\lambda)$ is determined by
Eq.(\ref{co:16}) in order to compare the values $\ln
\overline{a_n(\lambda)}$ and $\overline{\ln a_n(\lambda)}$.

Averaging $a_n(\lambda)$ over the random distributions of $\{1\}$ and
$\{0\}$ in the sequence $\{x^{(n)}\}$ (see (\ref{co:15})) and taking into
account the factorization of corresponding determinant (\ref{co:16}), we
get
$$
\overline{a_n(\lambda)}=\sum_{s=1}^{\infty}
\mathop{{\sum}'}_{\{l_1\ldots l_s\}} \mathop{{\sum}'}_{\{m_1\ldots m_s\}}
\prod_{j=1}^s\left[a_{l_j+1}(\lambda)\; (-\lambda)^{m_j-1}\right]
\eqno (B.1)
$$
where $\mathop{{\sum}'}$ means that summation runs over the sequences
$\{l_1\ldots l_s\}$ and $\{m_1\ldots m_s\}$ obeying the conditions:
$$
\begin{array}{l}
\disp \sum_{j=1}^s l_j=n\,(1-p) \medskip \\
\disp \sum_{j=1}^s m_j=n\,p
\end{array}
\eqno (B.2)
$$
Recall that $p$ is the fraction of zeros in the random matrix
$\widehat{M}^{\rm err}_n$ and according to Eq.(B.2) the
quantity ($n\,p$) is {\bf always integer}.

Using the integral representation of the Kronecker $\Delta$--function (see
Eq.(A.3)), we may rewrite Eq.(B.2) in the form of a grand canonical
distribution:
$$
\begin{array}{ccl}
\disp \overline{a_n(\lambda)}&=&
\disp \sum_{s=1}^{\infty} (-\lambda)^{np-s} \sum_{l_1\ldots l_s}
\prod_{j=1}^s a_{l_j+1}(\lambda)
\Delta\left[\sum_{j=1}^s l_j-n(1-p)\right] \medskip \\
&=&\disp (-\lambda)^{np}\frac{1}{2\pi i} \oint dz z^{-1+n(1-p)}
\sum_{s=1}^{\infty} \left[(-\lambda)^{-1} \sum_{l=0}^{\infty}
a_l(\lambda) z^{-l}\right]^s \medskip \\
&=& \disp (-\lambda)^{np} \frac{1}{2\pi i} \oint dz z^{-1+n(1-p)}
\frac{-b(\lambda,z)}{\lambda+b(\lambda,z)}
\end{array}
\eqno (B.3)
$$
where the function $b(\lambda,z)$ plays a role of generating function for
$a_l(\lambda)$:
$$
b(\lambda,z)=\sum_{l=1}^{\infty} a_l(\lambda) z^{-l}
$$

Recall that
$$
\begin{array}{l}
\disp a_l(\lambda)=\frac{2(-1)^n}{\sqrt{(3-\lambda)(1+\lambda)}}
(1+\lambda)^{l/2} \sin\left[(l+2)\theta\right] \medskip \\
\disp \sin\theta= \frac{1}{2}\sqrt{3-\lambda} \medskip \\
\disp \cos\theta= \frac{1}{2}\sqrt{1+\lambda}
\end{array}
\eqno (B.4)
$$

Collecting all equations together, we arrive at the following expression
$$
\overline{a_n(\lambda)}= (-\lambda)^{np} \frac{1}{2\pi i}
\oint dz z^{n(1-p)} \frac{\lambda (1+z^{-1})+z^{-1}}
{\lambda+\lambda^2z^{-1}+(\lambda^2-1)z^{-2}}
\eqno (B.5)
$$

Now we can use the definition of the Chebyshev's
polynomials generating function:
$$
\frac{1}{1-2tx+t^2}=\sum_{k=0}^{\infty}{\cal U}_k(x)t^k
$$
where
$$
{\cal U}_n(x)=\frac{\sin\left[(n+1)\arccos x\right]}{\sin[\arccos x]}
$$

After rather simple algebra we get the desired equation for the averaged
determinant:
$$
\overline{a_n(\lambda)}=(-1)^n\lambda^{np+1}
\left(\frac{\lambda^2-1}{\lambda}\right)^{\frac{n}{2}(1-p)}
\left[{\cal U}_{n(1-p)}(y)-\left(\frac{\lambda+1}
{\lambda(\lambda+1)}\right)^{1/2}{\cal U}_{n(1-p)-1}(y)\right]
\eqno (B.6)
$$
where
$$
y=\frac{\lambda^{3/2}}{2\sqrt{\lambda^2-1}}
\eqno (B.6a)
$$

As we can see, this result differs dramatically from the "quenched
case" (see Eq.(\ref{co:34})).

\newpage

\section*{Figure Captions}

\begin{fig}
Graphs, corresponding to: (a) free group $\Gamma_n$; (b) locally free
group ${\cal LF}_{n+1}$; (c) complete commutative group. In case of
locally free group the vertices $A$ and $B$ should be glued because they
represent one and the same word in group ${\cal LF}_{n+1}$.
\label{fig:1}
\end{fig}

\begin{fig}
Plot of the function $\ln S_n(\mu)$ in two regimes (Eq.(\ref{co:28a}))
\label{fig:2}
\end{fig}

\begin{fig}
Plot of the function $Q(p|\mu)$ for three fixed values of
$\mu=\{10,\,30,\,150\}$---see Eq.(\ref{co:plot}).
\label{fig:3}
\end{fig}

\begin{fig}
Schematic representation of the process associated with the "biased
Levy--flight" (BLF)--like random walk.
\label{fig:4}
\end{fig}

\end{document}